\documentclass[runningheads]{llncs}
\usepackage{graphicx}
\usepackage{amsmath}
\usepackage{subcaption}
\usepackage{hyperref}
\usepackage[strings]{underscore}

\begin{document}
\title{Assessing domain adaptation techniques for mitosis detection in multi-scanner breast cancer histopathology images}
\titlerunning{Domain Adaptation and Mitosis Detection}
%
\author{Jack Breen\inst{1}\orcidID{0000-0002-9020-3383} \and Kieran Zucker\inst{2}\orcidID{0000-0003-4385-3153} \and
Nicolas M. Orsi\inst{2}\orcidID{0000-0003-0890-0399} \and
Nishant Ravikumar\inst{1}\orcidID{0000-0003-0134-107X}}
\authorrunning{Breen et al.}
%
\institute{CISTIB Center for Computational Imaging and Simulation Technologies in Biomedicine, School of Computing, University of Leeds, UK \\
\email{\{scjjb,N.Ravikumar\}@leeds.ac.uk} \and
Leeds Institute of Medical Research at St James's, School of Medicine, University of Leeds, UK}
\maketitle              
\begin{abstract}
Breast cancer is the most commonly diagnosed cancer worldwide, with over two million new cases each year. During diagnostic tumour grading, pathologists manually count the number of dividing cells (mitotic figures) in biopsy or tumour resection specimens. Since the process is subjective and time-consuming, data-driven artificial intelligence (AI) methods have been developed to automatically detect mitotic figures. However, these methods often generalise poorly, with performance reduced by variations in tissue types, staining protocols, or the scanners used to digitise whole-slide images. Domain adaptation approaches have been adopted in various applications to mitigate this issue of domain shift. We evaluate two unsupervised domain adaptation methods, CycleGAN and Neural Style Transfer, using the MIDOG 2021 Challenge dataset. This challenge focuses on detecting mitotic figures in whole-slide images digitised using different scanners. Two baseline mitosis detection models based on U-Net and RetinaNet were investigated in combination with the aforementioned domain adaptation methods. Both baseline models achieved human expert level performance, but had reduced performance when evaluated on images which had been digitised using a different scanner. The domain adaptation techniques were each found to be beneficial for detection with data from some scanners but not for others, with the only average increase across all scanners being achieved by CycleGAN on the RetinaNet detector. These techniques require further refinement to ensure consistency in mitosis detection.
\end{abstract}

\section{Introduction}
Breast cancer is the most commonly diagnosed cancer worldwide, accounting for one quarter of all malignancies in women~\cite{Sung2021}. Diagnosis involves identifying cancer sub-type, grade, and molecular profile (oestrogen/progesterone receptor and HER2 amplification status). Counting dividing cells (mitotic figures) is a key task for pathologists within the Nottingham grading score, which combines mitotic count, tubule formation, and nuclear pleomorphism as a measure of the aggressiveness of the underlying malignancy. This grade carries prognostic information as part of the Nottingham Prognostic Index~\cite{WHO2019}.

Recent studies in human and veterinary pathology have highlighted that pathologists' detection of mitoses is both variable \cite{Bertram2021} and time-consuming~\cite{Laflamme2020}. This increases pressure on histopathology diagnostic services, where in the UK alone, only 3\% of departments are adequately staffed to meet diagnostic demand. As a result, 45\% of departments routinely outsource work and 50\% use locums~\cite{RCPath2018}, at significant cost. Recent advances in deep learning (DL)-driven automated techniques for mitosis detection have shown promise both for reducing inter-observer variability between pathologists, and for relieving some of the workload associated with the mitotic count.

Domain adaptation approaches in general aim to learn a mapping that reduces the gap between source and target data distributions. In the context of computer vision problems, they are employed to improve generalisation of image-based DL models to data from different domains during inference. While convolutional neural networks (CNNs) have proved a powerful tool for solving a multitude of vision problems, it is well established that they tend to over-fit to data in the training domain, and hence generalise poorly to target domains during inference. Domains in digital histopathology may include variations introduced from tissue staining processes, scanner properties, or the histological preparations being scanned. The domain adaptation methods we investigate focus on the visual appearance of an image, taking a content image and one or more style images, and creating a stylised representation of the content image. 

\subsection{MIDOG Challenge 2021 and Relevant Literature}

\begin{figure}[h]
  \centering
  \includegraphics[trim={0 0.3cm 0 0.15cm},width=0.99\linewidth]{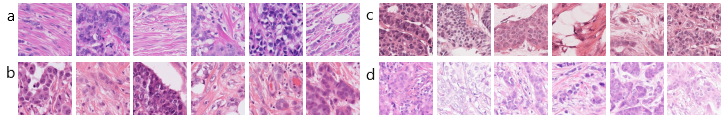}
  \caption{512x512 segments of MIDOG 2021 training data from (a) Hamamatsu XR (HXR), (b) Hamamatsu S360 (HS360), (c) Aperio CS (ACS), (d) Leica GT.}
  \label{UnlabelledCrops}
\end{figure}

The MIDOG Challenge was a competition on mitosis detection~\cite{MIDOG2021}, held at the International Conference on Medical Image Computing and Computer Assisted Intervention, 2021. The challenge provided 200 training images, each a 2mm$^2$ region of interest selected manually from haematoxylin and eosin (H\&E) stained breast cancer whole-slide images (WSIs). All samples were prepared using an identical staining process at the pathology laboratory of UMC Utrecht. These images were digitised using four different scanners: 50 each from Hamamatsu XR nanozoomer 2.0 (HXR), Hamamatsu S360 0.5 NA (HS360), Aperio ScanScope CS2 (ACS), and Leica GT450. All images digitised using the first three scanners were annotated via majority voting by three pathologists, to reduce inter-observer variability and single-observer bias. A deep learning model was used to suggest any potentially missed mitotic figures, which were also annotated by the pathologists. Scans from the Leica GT450 lacked annotations and were not used in this study. Algorithms submitted to the challenge were evaluated on a set of 80 WSIs from a combination of training set scanners and previously unseen scanners. 

Efforts were made to minimise selection bias during data collection for the challenge, by digitising all available breast cancer samples in the order that they had been sent to the pathology lab for examination. This led to a highly varied training set, with the number of annotations for a WSI ranging from 2 to 184. These annotations were sparsely distributed throughout the images, with most 512x512 crops containing no annotations at all. The difference between scanners is shown in Figure~\ref{UnlabelledCrops}. The scanners have different colour profiles, with ACS producing deeper red scans, and HXR producing scans with a blue-purple hue.

Prior to MIDOG~2021, few studies had investigated the impact of domain adaptation techniques on the detection of mitotic figures in images acquired using multiple scanners. A recent study used a domain adversarial neural network (DANN), a multi-task approach which combines domain adaptation and classification to learn image domain-independent representations~\cite{Otalora2019}. However, the method performed poorly when combined with stain normalisation. This suggests that the model learned to classify within specific domains rather than truly learning to classify in a domain-independent manner. 

Another study proposed a stain-transfer network using a GAN architecture with a U-Net encoder, adding an edge-weighted regularisation to retain basic structures from the input images~\cite{BenTaieb2018}. This was found to improve performance for patch-level classification, but was not tested on pixel-level predictions for localising and counting mitoses. In histopathology more generally, researchers have evaluated the benefit of domain adaptation for segmentation~\cite{Khan2014}, detection~\cite{Liimatainen2019}, and classification tasks \cite{Shin2021}. 

The most common techniques are stain normalisations based upon matching colour distribution, and GANs for holistic domain adaptation. The domain adaptation methods we investigate in the current study have a limited amount of previous research in histopathology. Two studies investigated Neural Style Transfer (NST) for transferring stains across cell-level images, but without evaluating the subsequent effects on a computer vision model \cite{Ganesh2019,Izadyyazdanabadi2019}. To the best of our knowledge, these are the only studies to use this method for histopathology without significant adjustments, such as changing to an adversarial loss function~\cite{Nishar2020}. The original NST method has not been evaluated as a domain adaptation tool for cell-level classification, segmentation, or detection. CycleGAN is increasingly popular in histopathology \cite{Shin2021}, and while previous research in mitosis detection has evaluated it for switching between stains \cite{Mercan2020}, it has not previously been evaluated for domain adaptation within the same staining process. 

\section{Methods}
\subsection{Mitosis Detection Models}
\begin{figure}[h]
\begin{center}
    \subfloat[]{{\includegraphics[trim={1cm 0.9cm 1cm 0cm},width=2.5cm]{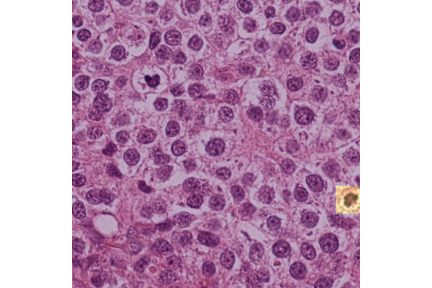} }}
    \subfloat[]{{\includegraphics[trim={1cm 0.9cm 1cm 0cm},width=2.5cm]{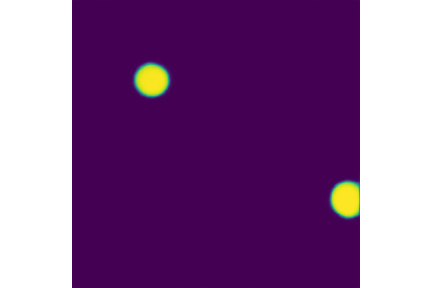} }}%
    \subfloat[]{{\includegraphics[trim={1cm 0.9cm 1cm 0cm},width=2.5cm]{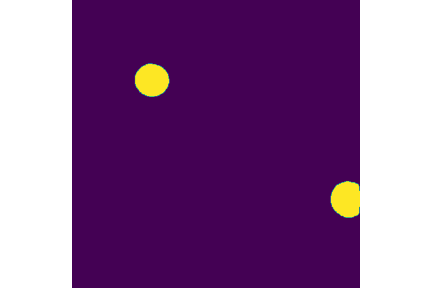} }}%
    \subfloat[]{{\includegraphics[trim={1cm 0.9cm 1cm 0cm},width=2.5cm]{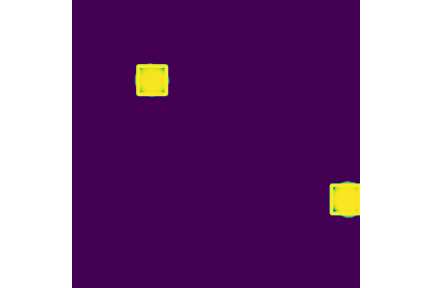} }}%
    \caption{U-Net post-processing procedure. (a) ground-truth image with one mitotic figure (b) predictions map output from U-Net (c) binary map (d) bounding box predictions. This example has one True Positive prediction (upper left) and one False Positive prediction (lower right).}
    \label{U-NetPreds}%
\end{center}    
\end{figure}

\textbf{U-Net} is an architecture for semantic segmentation which combines multiple layers of downsampling to generate a multi-scale feature mapping~\cite{Ronneberger2015}. This method requires segmentation masks for training, which we generated by taking each pixel to be a 1 if it was within a mitotic figure bounding box and a 0 otherwise. 
U-Net outputs a probability map, which we converted to bounding box predictions through a multi-step process, shown in Figure \ref{U-NetPreds}. First a binary map was generated by applying a threshold to the probability map. Objects were subsequently extracted from this by selecting external contours. Any detection with a height or width less than 10 pixels was assumed to be an artifact and was removed, as this was empirically found to improve robustness. The remaining detections had a bounding box placed around their center at the same size as the original annotations. 
We used a combination of binary cross entropy loss, dice loss and focal loss, which we weighted heavily towards the focal loss as this performs well on unbalanced datasets~\cite{Lin2020}.

\textbf{RetinaNet} is a one-stage detection algorithm which feeds inputs through an encoder and a feature pyramid network to generate multi-scale features. These features are used for simultaneous bounding box regression and classification~\cite{Lin2020}. This has previously been shown to perform at the level of an expert pathologist for quantifying pulmonary haemosiderophages~\cite{Marzahl2020}.

\subsection{Domain Adaptation Methods}
\textbf{Neural Style Transfer} (NST) is a one-to-one domain adaptation method, casting the style of one image on to the content of another~\cite{Gatys2016}. NST uses intermediate layers of a pre-trained VGG19 CNN to extract features representing the style and content of each input image. The loss function combines a style loss, which quantifies the difference between the output image and the input style image, and a content loss, which quantifies the difference between the output and the input content images. 

\textbf{CycleGAN} is a generative adversarial network used for visual domain adaptation~\cite{Zhu2017}. CycleGAN uses two GANs, one to produce a stylised image and another to recreate the original input image from the stylised image. This attempts to overcome mode collapse, where a generator creates the same output regardless of input. The performance at reproducing the original input image is measured by a cycle consistency loss, and the performance at transferring the style is measured by an adversarial loss. As GAN losses tend not to give a clear indication of convergence, we use the Fréchet Inception Distance to decide when to stop training \cite{Heusel2017}. 

\textbf{Macenko normalisation} is a common normalisation approach in H\&E-stained images which accounts for each stain separately \cite{Macenko2009}. This was used for comparison to the domain adaptation approaches. The normalisation was applied to both the training data and the evaluation data, where the domain adaptation approaches were only applied to evaluation data.

\begin{figure}[h]
  \centering
  \includegraphics[width=0.95\linewidth]{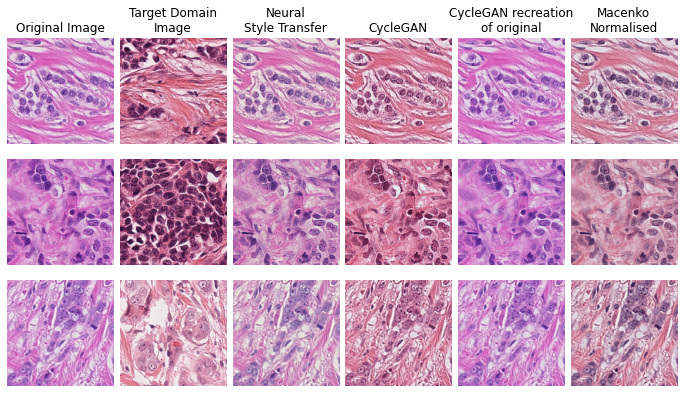}
  \caption{Domain Adaptation and normalisation approaches applied to three crops from the Hamamatsu S360 scanner, with Aperio CS as the target domain.}
  \label{StylesFig}
\end{figure}

Examples of all three methods are shown in Figure \ref{StylesFig}. Neural Style Transfer changes the colour profile of scans much less than CycleGAN, as it was found that running NST for more iterations led to very poor mitosis detection performance, as the resulting images were too artificial. 

\subsection{Implementation} 
Models were evaluated using the cell-wise F1 score, which is the harmonic mean of the precision and recall. This metric punishes false detections and missed mitotic figures equally. True positives are defined as predictions made within 30 pixels of the center of an annotated mitotic figure. Both domain adaptation methods were used as normalisation approaches during inference, not during training. This approach keeps the detection models completely agnostic to the testing domain, ensuring that any difference to detection performance is a result of the domain adaptation approaches alone.

All experiments were undertaken with a single GPU on Google Colab. Both of the detectors were implemented with a ResNet encoder pre-trained on ImageNet, with ResNet101 found to be optimal for RetinaNet and ResNet152 for U-Net. We evaluated our methods using a three-fold approach where two annotated scanners were used for training and the other one was withheld for evaluation as an unseen domain. The average F1 score and 95\% confidence intervals were calculated using 10,000 epoch bootstrapping, with evaluations on non-overlapping 512x512 crops from all available WSIs in the training domain (40 training WSIs and 10 test WSIs per scanner), and from all 50 WSIs in the external domain.

\section{Results}
\begin{figure}[h]
  \centering
  \includegraphics[width=0.95\linewidth]{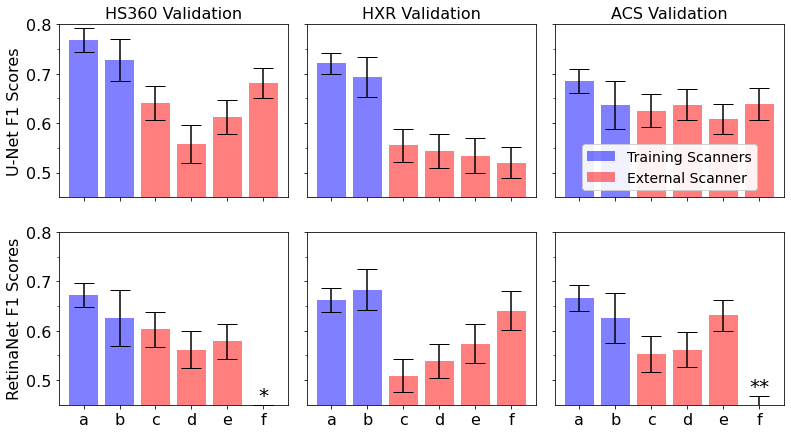}
  \caption{F1 scores from each detection model, where two scanners were used for training and one for validation, on (a) the training set (b) a test set from the training domain (c) the external domain (d) the external domain with NST (e)~the external domain with CycleGAN (f) the external domain with Macenko. 95\% confidence intervals shown in black. $\text{*}0.243\pm0.043$. $\text{**}0.426\pm0.041$.}
  \label{Results}
\end{figure}

Mitosis detection results from all experiments conducted in this study are summarised as bar plots of F1 scores in Figure \ref{Results} with 95\% confidence intervals. The best average performance on unseen data was achieved by the U-Net with Macenko normalisation, with the baseline U-Net and U-Net+CycleGAN close behind. At the time of MIDOG~2021, our best results were from U-Net without domain adaptation, so we trained this model on data from all three training scanners. Our model achieved an F1 score of 0.693 on the preliminary test set, and 0.686 on the final test set. Precision and recall were very similar, at 0.686 and 0.685, respectively, indicating that the model was balanced.

\section{Discussion}
The performance of both baseline detection algorithms is found to be comparable to human performance, with an average F1 score of 0.69 for U-Net and 0.65 for RetinaNet, compared to the human score of 0.68 in a comparable study \cite{Bertram2021}. The average performance on the unseen domain is lower, at 0.61 and 0.56, respectively. Detection performance was generally lower when the Hamamatsu XR was used as an unseen scanner, which is likely caused by the scanner having a significantly different colour profile to the other two scanners. Both domain adaptation methods slightly improved performance on the RetinaNet, though this performance was still lower than on the other scanners.

Overall, the domain adaptation methods performed inconsistently, with CycleGAN improving the average F1 score for RetinaNet but not for U-Net, and NST improving the average score for one of the three unseen domains for each detector, but degrading average performance. Macenko normalisation gave a slight increase in average detection performance for the U-Net, but was very inconsistent for RetinaNet, with both domain adaptation techniques performing better on average. Due to computational limits and long run-times for the domain adaptation methods, hyperparameter optimization was limited. Our hyperparameters were thus influenced by similar works and by practicalities, and were likely to be sub-optimal for mitotic figure detection, which was not investigated in these similar works. 

Future work should focus on evaluating the source of inconsistencies in domain adaptation methods applied to digital histopathology applications, to improve reliability. This may include combining images from different scanners to create a more general target domain for CycleGAN. Furthermore, the effects of training both detection models with artificial images generated by the domain adaptation methods should be evaluated as this may make the detectors more robust to domain shifts. To better evaluate the domain adaptation methods, it would be beneficial to also compare their respective run-times. 

\section{Conclusion}
We implemented\footnote{\url{https://github.com/scjjb/MIDOG_Domain_Adaptation}} two unsupervised domain adaptation techniques, CycleGAN and Neural Style Transfer, for overcoming scanner-driven domain shifts in histology images, and enabling robust mitosis detection. These were applied to transform test data acquired using scanners previously unseen by the detection models. Our baseline detection methods, U-Net and RetinaNet, performed comparably to a human expert on data from the training domain, with reduced performance on data from an unseen domain. Both domain adaptation techniques were found to improve detection performance for some of the unseen domains but not for all. Average detection performance across all unseen domains was only improved using CycleGAN in combination with the RetinaNet detector. This partial success justifies the need for further investigation to understand and overcome these inconsistencies. Improved techniques for modelling in the presence of domain-shifts, or for learning domain-invariant features, will be essential for accelerating the deployment and adoption of automated tools in routine diagnostic practice.

%
%
%
\bibliographystyle{splncs04}
\bibliography{library}

\end{document}